# CABLE NETWORK MANAGEMENT INFRASTRUCTURE EVOLUTION

L. Alberto Campos, Jennifer Andreoli-Fang and Vivek Ganti
Cable Television Laboratories

*Abstract*

*An approach to enable advanced troubleshooting, granular analysis and service quality of experience assessment is presented. The use of topology information in the identification of each cable network element along with granular information of the element configuration and health is proposed. This technique covers multiple layers including the service layer. At the physical layers street addresses, taps and amplifiers are used to identify impairment location. All layers are leveraged to measure network and service reliability, service degradation and to quantify quality of experience. A cable infrastructure implementation is described as an example.*

## BACKGROUND

Historically cable networks consisted of a multitude of services that developed relatively independent from each other. There have been always considerations for coexistence since the services are sharing the same physical environment but their designs had limited relation. Analog video services initiated as a basic transport of broadcast signals over the cable infrastructure. The introduction of digital video had only to be cognizant of the legacy analog video channelization in addition to fidelity requirements for suitable coexistence. DOCSIS® went a step further as in addition to adopting the channelization of legacy services, DOCSIS® also adopted the same downstream physical layer as the digital video downstream. This provided synergies that led to common downstream transmission systems called Edge-QAMs and the modular systems that are described in the M-CMTS specifications. The HFC telephony systems have been for the most part an adaptation of circuit switched telephony systems to the cable environment. The limited relation of these service platforms has also resulted in separated isolated management systems with different performance indicators and management parameters. The industry has yet not gone through the trouble of examining the relation of all these indicators and parameters to realize potential simplifications in network management systems. Recently with the advent of DOCSIS® 3.1 and the implied migration to a common IP platform for all services, there is the potential of introducing a unified set of performance transport metrics and performance thresholds with significant enhancements regarding how the networks are managed and provisioned.

Moreover, the capabilities of the tools to handle big data systems and perform complex analysis on them have significantly improved in recent years. In the cable environment we have been introducing more elaborate network maintenance techniques such as Proactive Network Maintenance, which leverage granular plant health probing and data correlation and analysis.

This paper proposes a strategy for advanced network monitoring and management to support end-to-end service analysis, on an individual subscriber service basis. This leverages the collection of data in a granular fashion from all cable monitoring and management systems. A key goal of this approach is to develop a management system capable of assessing end-to-end service performance, quantifying quality of experience, determining network health, network reliability and determining the resources to provide services. The latter is also useful to assess the resources remaining available to provide other services. Provisioning of business services in general and backhaul services in particular are use case scenarios of particular interest.

## LAYERED APPROACH EVALUATION

In order to assess end-to-end service performance, the assessment will have to encompass/traverse multiple layers. The network management systems are traditionally specific to a section of the network and to a layer of the network. These management systems are typically isolated from each other. There is a need to extract information from all these layers and correlate such information to assess end-to-end service performance. Figure 1 shows a schematic representation of the typical components in a cable network.

Figure 1 shows a collection of components that are used to provide the services that are typically delivered through a cable TV network. By no means one unified management system is used to manage the components represented in Figure 1. Yet, a single service traverses different sections of this network that is the under control of different management systems.

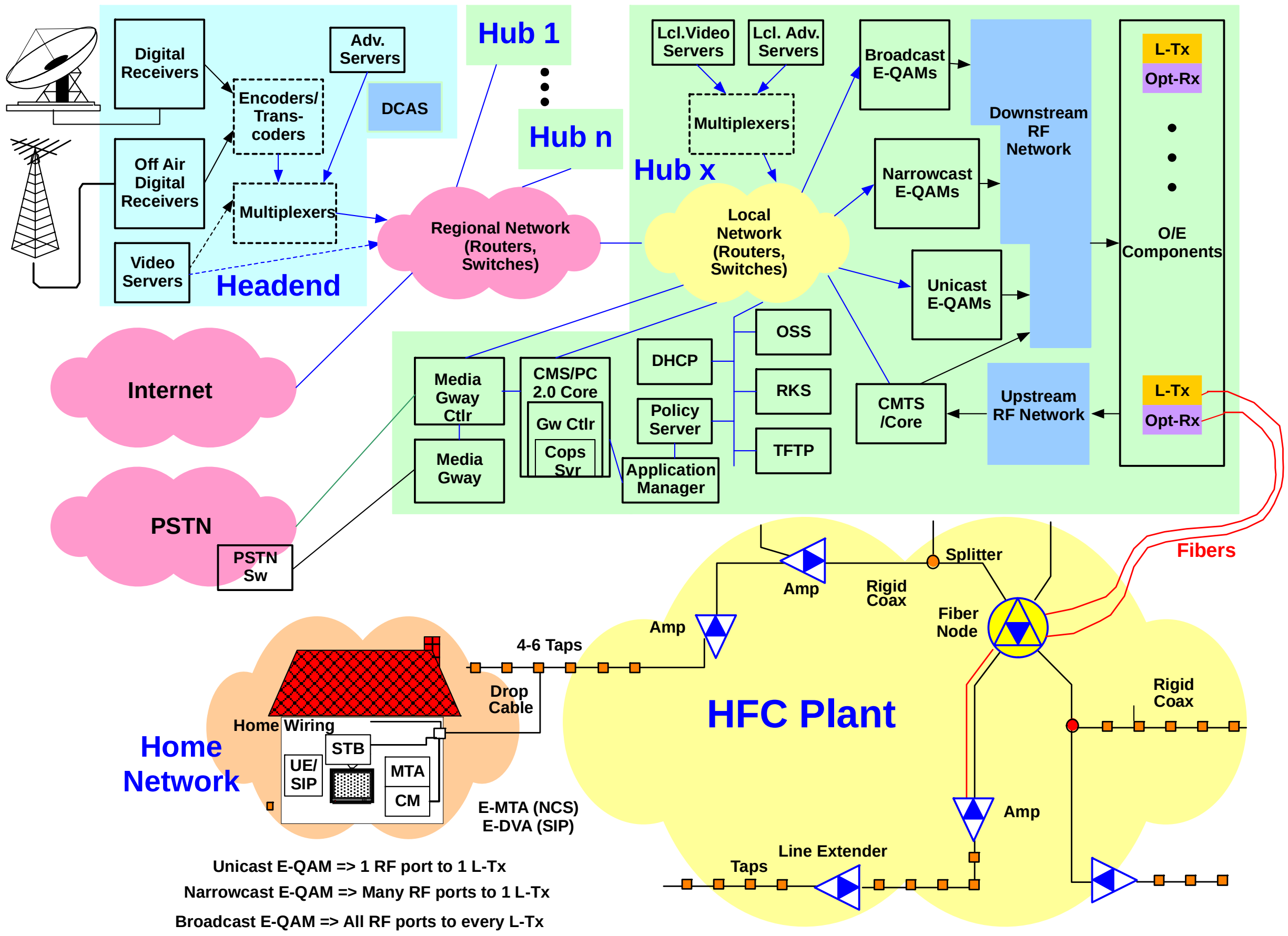

Figure 1 – Cable Network Elements for Service Delivery

In order to analyze an end-to-end service, one has to select the subset of the network components that participate in the delivery of that service. This subset of the network has a corresponding set of management systems that for end-to-end service delivery would have to be coordinated in a cohesive fashion. To facilitate analysis, one can further divide this service into sub-services. For example, a further segmentation for voice services would be call-signaling and call-payload transport. A segmentation for data sub-services could be data-service provisioning, data-authentication and data-payload transport.

Figure 2 shows a subset of the network containing the end-to-end paths of a data-service payload transport example and a video service payload transport example. These end-to-end service paths are also divided based on the layers they traverse. For practical analysis purposes we divide these service paths into four layers. The physical (PHY) layer, the link or medium access control (MAC) layer, the TCP/IP layer and the services and applications layer. As shown in Figure 2 there are different network elements that participate depending on the layer.

In the data-payload transport service example the PC and the data server (in light blue) are the elements that participate in the services and applications layer while for the video-payload transport example a video-terminal and a video server (in red) are the elements that participate in the services and applications layer.

Examining the MAC layer participant elements in the data-payload transport service example, the elements involved are:
PC, cable modem, Edge-QAMs, CMTS core, local network, regional network, public internet and data server.

In the case of the PHY layer participants of the data payload transport service example, the elements involved are:
PC, cable modem, taps, amplifiers, splitter, optical node, laser transmitter and optoelectronic receiver, downstream and upstream RF networks, Edge-QAMs and CMTS core, local network, regional network, public internet and data server.

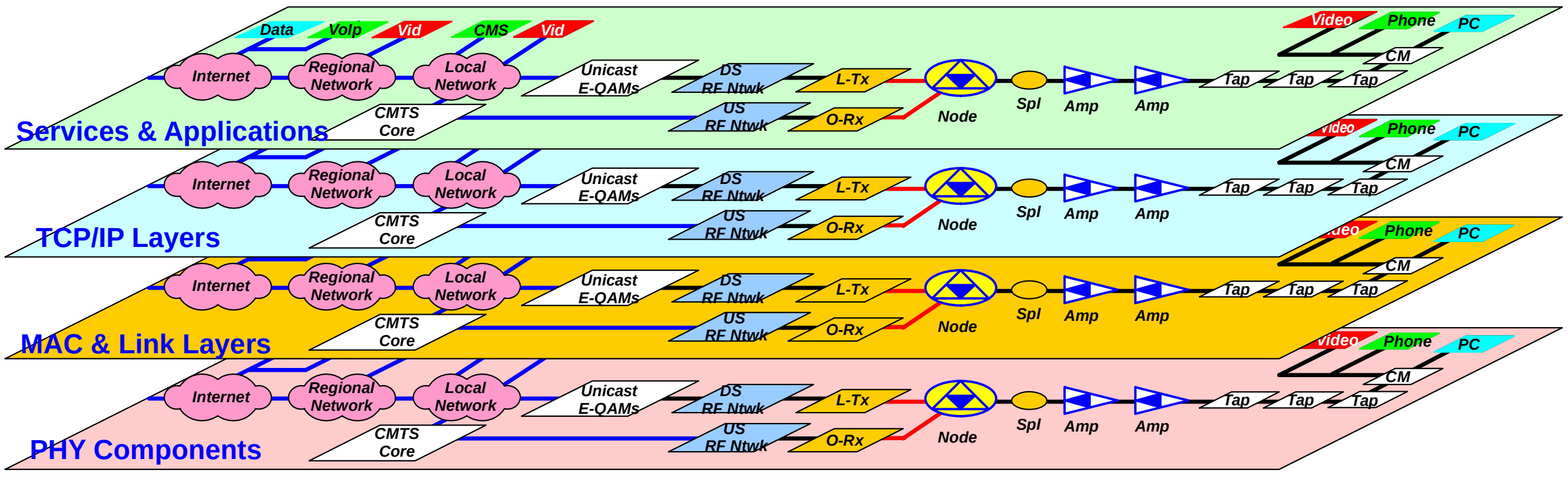

Figure 2 – Cable Network Subset Dissected in Layers

Based on the available tools that can troubleshoot issues at different layers, different elements get involved in the analysis process. There are elements that participate in more than one layer. These are elements that can be leveraged as anchor-elements which enable the correlation of data across multiple layers.

There are many systems that currently participate in the analysis of service performance. Each system typically has its own performance metrics. Many of these metrics are service layer metrics, symptoms gathered in these upper layer metrics are also used for issues that may be happening in the lower layers. Accessing directly the appropriate lower layer would be more efficient but many times the management system of the upper layer service doesn't have access to the lower layer. Breaking the silos of management systems and coordinating the management information enable the implementation of end-to-end service management.

Network Connectivity Relationship Among CATV Elements

The way management systems are used, highlights the importance of establishing element connectivity relations and understanding in which layer this connectivity relation takes place.

In the HFC portion of the network, the traffic follows the general paths from the Headend or Hub to the end device. In this portion of the network, which follows a tree and branch topology, there is a simple way to describe the connectivity relations using the "Long Name convention" which was proposed in [1].

This long name convention for PHY Layer uses the device naming technique that describes the path between the device or component starting from the element where the service originates to where it ends. In the case where there is a common point for all elements using that layer, a common northbound device that is uniquely described can be used. In the case of an upstream RF transmission to the CMTS, the optical node can be used as the common, uniquely identified northbound device where the tree and branch topology converges like the optical node (Figure 3). In this notation, subscripts are used to indicate branching out of multi-port devices.

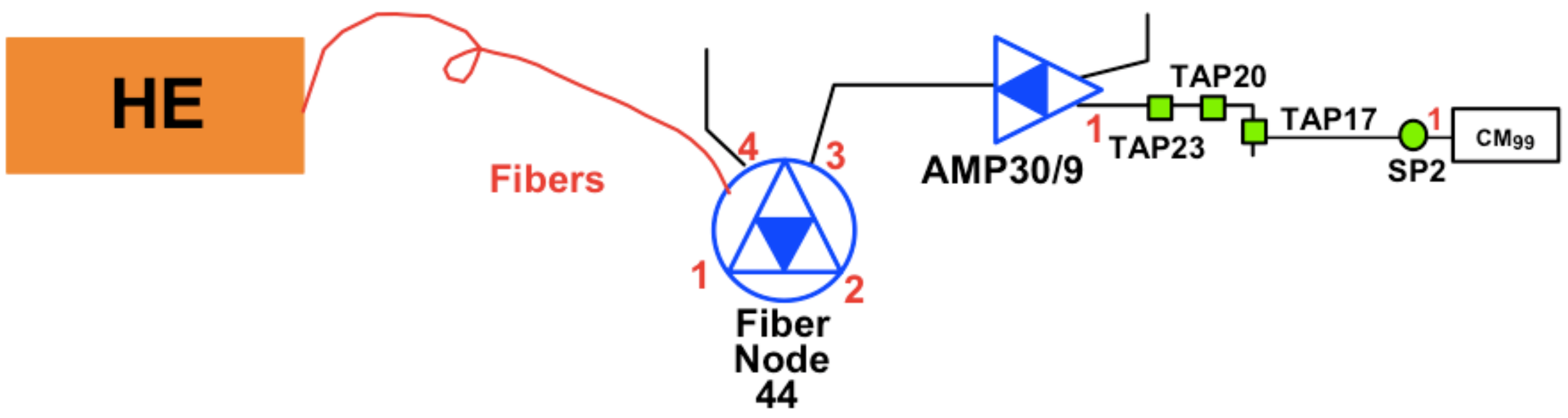

Figure 3 – PHY Layer Long Name Convention

In this case the cable modem with unique name CM99 would also be known in its long name convention as:

$FN44_3$-AMP30/91-TAP23-TAP20-$TAP17_1$-$SP2_1$-CM99

and the 23dB Tap can be identified by it's long name convention;

$FN44_3$-AMP30/91-TAP23

This naming convention representation is suitable for the PHY layer that follows a tree and branch topology. The naming convention is also suitable for a sequential path of components being traverse and for point-to-point topologies. It is however, not suitable to capture all possible connection and routes for mesh topologies enabled in the MAC and TCP/IP layers, but it is suitable for tracing the communications path for a particular service or application through the TCP/IP, MAC and PHY layers. This occurs as a direct consequence of the IP routing principle that avoids routing loops.

Therefore for describing mesh topologies in the MAC and IP layers, the entire connectivity relationships are required. The level of complexity increases since the connectivity relationships must be kept for all layers. The advancements in computational power and analysis capabilities that come along with big data, have made feasible tackling problems with this level of complexity. In the case of service layer analysis for a particular service rendered to a device, a long name convention can also be used. A device subscribed to a service or sub-service may have a long name in the IP layer another one in the MAC layer and a third one in the PHY layer. For example the naming convention in the MAC layer for a Video Download Service to a video terminal with an integrated STB is;

LocalVideoSvr33- $EthSwitch17_{15}$-$EQAM2_{12}$-VideoTerm99

<u>Practical Approach to Granular Service Performance Measurements</u>

It is many times easy to leverage the ever increasing computational power. Nevertheless this carries also complexity of implementation and in operations. It is desirable to strike a balance between granularity and simplicity.

In the multilayer management environment we have in our systems and services, there are significant dependencies between the different performance metrics and thresholds. As the management systems become unified through either interactivity among them or through a master management system that is able to access them, significant simplification on the way we determine health and performance of our networks becomes feasible. As a result of this unified management approach, the silos between the performance metrics databases originated from separate management systems are broken down.

If a metric used in a system is dependent on a root metric, then only the root metric should be used. A review of all performance metrics across the different management platforms is required. The abundant metrics currently available on the different systems lead to operational expenses that could be avoided through a network management optimization process.

There is a multitude of performance metrics that could be considered when evaluating broadband services and applications [2]. Figure 4 shows an example of broadband services and applications and their most relevant performance metrics.

| Services & Applications | General Metrics | | | | | | Service Specific Metrics | | | | | | |
|---|---|---|---|---|---|---|---|---|---|---|---|---|---|
| | Instant Availability | Packet Loss | Latency | Jitter | Peak Rate | Sustainable Rate | Controllability | Avail. Content | Switching Speed | Capacity | Feature Richness | Drop Call % | Competence |
| Channel Surfing | | | H | | | | | | H | | | | |
| Data BE | | | | | H | | | | | | | | |
| Video Conferencing | | | H | H | | H | | | | | | | |
| VoD | | H | | | | H | | H | | | | | |
| Music Services | | H | | | | H | | H | | | | | |
| Gaming Client | | H | H | | | | | | | | | | |
| Gaming Server | | H | H | | | | | | | | | | |
| SuperDownloading | | | | | H | H | | | | | | | |
| SuperUploading | | | | | H | H | | | | | | | |
| Storage | | | | | | | | | | H | | | |
| Tele-Medicine | H | H | H | | H | H | | | | H | | | |
| Tele-Education | | H | H | | | H | | | | | | | |
| Tele-Commuting | | | | | | H | | | | | | | |
| Tele-Library | | | | | H | | | H | | H | | | |
| Customer Support | | | | | | | | | | | | | H |
| Home Security | H | | | | | | | | | | H | | |
| Home Remote Ctrl | H | | | | | | H | | | | H | | |
| Telephony | | | H | H | | | | | | | H | H | |
| Video SD | | H | | | | H | | H | | | | | |
| Video HD | | H | | | | H | | H | | | | | |
| Video Low Res | | H | | | | H | | H | | | | | |
| Business Services | H | H | H | H | H | H | | | | | | | |

Figure 4 - Multiple Performance Metrics used for Different Services

In Figure 4, the red cell with an H indicates high relevance of the performance metric for that particular service or application.

Collapsing Multiple Performance Metrics to Common Performance Index

As mentioned above there is an advantage in reducing the number of performance metrics. If a relationship between the performance metrics can be determined, this relationship can be leverage to reduce the number of performance metrics. All possible performance metric relationships ought to be explored. There may be relationships between percentage voice call-drops and packet loss or relationship between video macro-blocking and packet loss. If a strong relation is determine only one parameter can be used.

In the DOCSIS® 2.0 and 3.0 HFC environment for example, there is an intuitive correlation between packet loss, latency and jitter performance. These are fundamental objective network performance metrics through which other upper layer performance metrics may be derived. We proceed to explore the relationship of these metrics.

A DOCSIS® 3.0 network consisting of a CMTS and CMs is setup. In this network loading of the channel is performed and the latency, jitter and packet are measured. Figure 5 shows the behavior of these metrics with channel loading.

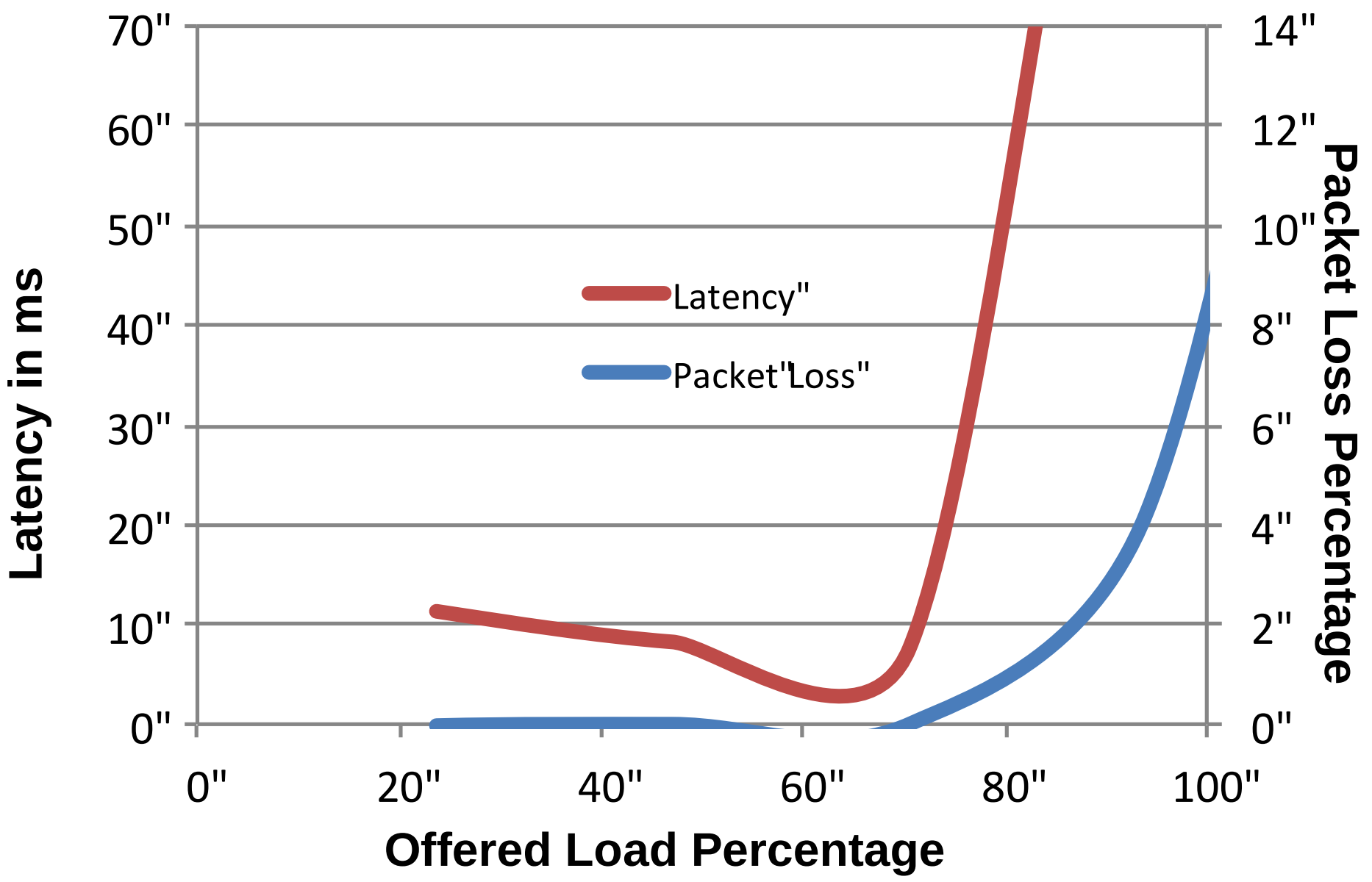

Figure 5. Correlation of DOCSIS Performance Metrics

Each curve correspond to a particular packet size, an additional curve based on an average packet size distribution can be generated. The resulting graphs indicate that within the DOCSIS® environment there is a strong correlation between packet loss, jitter and calibrated latency. Such a relationship enables the use of a unified performance index by collapsing multiple performance metrics into a common performance index.

For example, the main performance metric for gaming is latency. Let's assume that the threshold for unacceptable latency in gaming is 20 msec. Let's also assume that the main performance metric for video streaming is packet loss and the threshold for packet loss is $10^{-6}$ PER.

The sample performance table below shows the relation between packet loss, jitter and latency for a specific DOCSIS® network.

| Performance Index | Packet Lloss % Lost/Sent Ratio | Latency ms |
|---|---|---|
| 1 | 0.01 | 9 |
| 2 | 0.2 | 10 |
| 3 | 0.5 | 20 |
| 4 | 1 | 40 |
| 5 | 2 | 80 |

Table 1 - Performance Index Equivalency Chart

Having a smaller set of performance metrics defined the performance of all broadband services is very attractive. This could limit complexity in assessing performance on a per service basis, yet enable the granularity of performance on a per user per service basis.

## GRANULAR DATA MODEL DEVELOPMENT AND AUTOMATION OF ANALYSIS

Some of attributes foreseen for future network management systems are those of higher granularity and capable of end-to-end assessments. In addition to these or to enable these, it is proposed here that the future network management systems also have to be aware of the network connectivity relations and the network layers the different management systems influence.

The higher granularity has to be assumed in two dimensions. One dimension relates to the increase granularity in element coverage (a larger set of components enter the management systems). A second dimension relates to the increased level of detail in which all the components are managed. As mentioned above it is beneficial to introduce network layer awareness and connectivity awareness. Most of the problems in the network even though they may be symptomatic across multiple layers, their cause is layer specific. Network layer awareness allows better assessment of network health and better troubleshooting of problems. In the network connectivity is a concept that is layer specific. What is view as connected in the physical layer is not necessarily view as connected in the TCP/IP layer or in the services and applications layer (Figure 2). Connectivity awareness in the databases containing the elements allows among many other things for automation of service reliability estimation, impairment localization, network performance assessment, capacity planning etc.

A couple of examples of data model elements that fit the desired goals of granularity, connectivity awareness and network layer awareness are included. The goal is to have this data model description of all of our network elements to fulfill the desired capability for end-to-end analysis.

Figure 6 shows the physical layer portion of the CMTS element model that has the characteristics proposed in this approach. It is intended to have a similar description for all network elements that participate in the physical layer. Figure 1 and Figure 2 illustrate the diversity of components that are encompassed in the granular database of cable network elements.

An important component of the data model is the downtime counter which is used to assess reliability. Another key component of this proposed model is the connectivity relation for each interface. This, in the case of the physical layer, would allow the manipulation of elements to automatically determine location of impairment or reliability. A third fundamental ingredient of the data model is the use of latitude and longitude or other means to determine location. This model uses location based on the type of element. There are three element types identified in the data model. One is a single location type, such as node or termination. A second is a linear type to represent as a linear device like a cable that has two vertices an location information for two points. The third is polygon type to represent a device covering an area such as a network or a linear multipoint element such as a cable conduit that requires a path to be specified.

To facilitate management there is some grouping proposed in this model. These are done through the association to a node, to a hub and to a system. The above is a sample approach not intended to be all inclusive but to demonstrate the required structure and to be expandable following the same structure.

| | | | |
|---|---|---|---|
| Element | | CMTS | |
| ElementID | | CMTS4 | Unique Within Fiber Node |
| P. Fiber Node Assoc. | | NA | Unique within Hub |
| P. Hub Assoc. | | Hb10 | Unique within System |
| P. System Assoc. | | S5 | Unique |
| Name | | ArrisC4-2 | |
| Element Type | | Node | Polygon/Cloud Node Link Termination |
| Location Vrt 1 | | Lat-Lon | |
| Layer 1 | | Physical | |
| Physical Element Type | | Node | |
| Number of Interfaces | | 9 | (ports) |
| | InterfaceID | 10 | Unique within Element |
| | Vrt Assoc Connection | 1 | 1 Vrt/Node, 2 Vrt/Link, Multiple Vrt/Polygon or Cloud |
| | Interface Rank | Primary | Primary/Secondary |
| | Interface Category | Fiber | Coax/Fiber/CatN/Wireless |
| | Interface Type | FC/PC | FC/PC, APC, JS |
| | Wavelength(s) | 1550 | nm |
| | Power Level(s) | 10 | dBm |
| | Tx/Rx Mode | Full Duplex | Tx Only, Rx Only, Full Duplex, Half Duplex |
| | Downtime Counter | | sec |
| | InterfaceID | 43 | Unique within Element |
| | Vrt Assoc Connection | 1 | 1 Vrt/Node, 2 Vrt/Link, Multiple Vrt/Polygon or Cloud |
| | Interface Rank | Secondary | Primary/Secondary |
| | Interface Category | Coax | Coax/Fiber/CatN/Wireless |
| | Interface Type | F | F/KS |
| | Level wrt Primary Fwd | NA | dB |
| | Level wrt Primary Rev | NA | dB |
| | Tx/Rx Mode | Rx Only | Tx Only, Rx Only, Full Duplex, Half Duplex |
| | Downtime Counter | | sec |
| Intended Paths | | 10-11, 10-20, 11-20, 11-10, 20-10, 20-11, 40-10, 40-11, 40-20, 41-10, 41-11, 41-20, 42-10, 42-11, 42-20, 43-10, 43-11, 43-20 | |

Figure 6 - Layer 1 Granular Data Representation for CMTS Data Element Model

Figure 7 shows the MAC layer and the TCP/IP layers portion of the CMTS element model that has the characteristics for this proposed data model. For the purposes of illustration only a few of the many interfaces in the CMTS data element model are shown. It is worth noting that each layer has connectivity information that is likely different the other layers of the same element. This is highlighted in Figure 2 as the participation of elements in the different layers is different and the resulting connectivity within a layer varies.

Elements such as a splitter, a tap, a fiber node, a coaxial cable segment participate are only described in the PHY layer, while elements like a cable modem or an Ethernet switch are described in the PHY layer and the MAC layer. An element like a CMTS or a home WiFi router are described in the TCP/IP layer in addition to the lower layers and elements like video servers, an EMTA or a PC are described using all layers.

| | | | |
|---|---|---|---|
| Layer 2 | MAC | | |
| MAC Element Type | Node | | |
| Number of Interfaces | | 9 | (ports) |
| | MAC InterfaceID | 1 | Unique within Element |
| | Phy IF Assoc | 10 | |
| | Connection | | |
| | Interface Rank | Primary | Primary/Secondary |
| | Interface Category | Eth | EPON, GPON, Eth, DOCSIS, 802.11, LTE |
| | Interface Type | 10G-Eth | 10BT, 100BT, 1G-Eth,10G-Eth,100G-Eth, |
| | MAC Address | | |
| | Downtime Counter | | sec |
| | MAC InterfaceID | 6 | Unique within Element |
| | Phy IF Assoc | 43 | |
| | Connection | | |
| | Interface Rank | Secondary | Primary/Secondary |
| | Interface Category | DOCSIS | EPON, GPON, Eth, DOCSIS, 802.11, LTE |
| | Interface Type | D3.0 | D1.0, D1.1, D2.0, D3.0, D3.1 |
| | MAC Address | | |
| | Downtime Counter | | sec |
| Intended Paths | | Any-to-Any | |
| Layer 3 | IP | | |
| L3 Element Type | Node | | |
| Number of L3 Interfaces | | 12 | |
| | L3 InterfaceID | 1 | Unique within Element |
| | MAC IF Assoc | 1 | |
| | Connection ??? | ??? | |
| | Interface Rank | Secondary | Primary/Secondary |
| | Interface Category | IP | IP, other |
| | Interface Type | IPv4 | IPv4, IPv6 |
| | IP Address | | |
| | Subnet | | |
| | Class | | |
| | Gateway | | |
| | Downtime Counter | | sec |
| | L3 InterfaceID | 12 | Unique within Element |
| | MAC IF Assoc | 1 | |
| | Connection ??? | ??? | |
| | Interface Rank | Secondary | Primary/Secondary |
| | Interface Category | IP | IP, other |
| | Interface Type | IPv4 | IPv4, IPv6 |
| | IP Address | | |
| | Subnet | | |
| | Class | | |
| | Gateway | | |
| | Downtime Counter | | sec |

Figure 7 - Layer 2 and 3 Granular Data Representation for CMTS Data Element Model

The data model just described has the characteristics of being; granular, layered, connection oriented and end-to-end analysis friendly. Next we exploit the use of the database and the relationships among its elements to implement in advanced network management and monitoring tasks.

USE CASE SCENARIOS LEVERAGING NETWORK MANAGEMENT APPROACH

Several use case scenarios of advanced network maintenance and management exercises that leverage the proposed data model are described.

First Use Case Scenario – Data Network Performance Analysis

In the sample scenario where gaming and video streaming bundled services are simultaneously supported, the minimum performance index supporting both services is performance index 4. In a group of devices that are evaluated the performance index varies around a mean. The mean is calculated from the group of devices that share certain topological characteristics (a node, MAC domain, a branch within the coaxial network etc.)

Figure 8 shows a sample data collection of different serving groups corresponding to multiple nodes and the multiple DOCSIS MAC layer domains under evaluation. The measured performance of end devices from different serving groups are distributed across a performance index range. The color coding is such that if two sigmas (standard deviations) of the population are below the performance threshold it is red. If two sigmas are above the performance threshold is green and yellow in other cases.

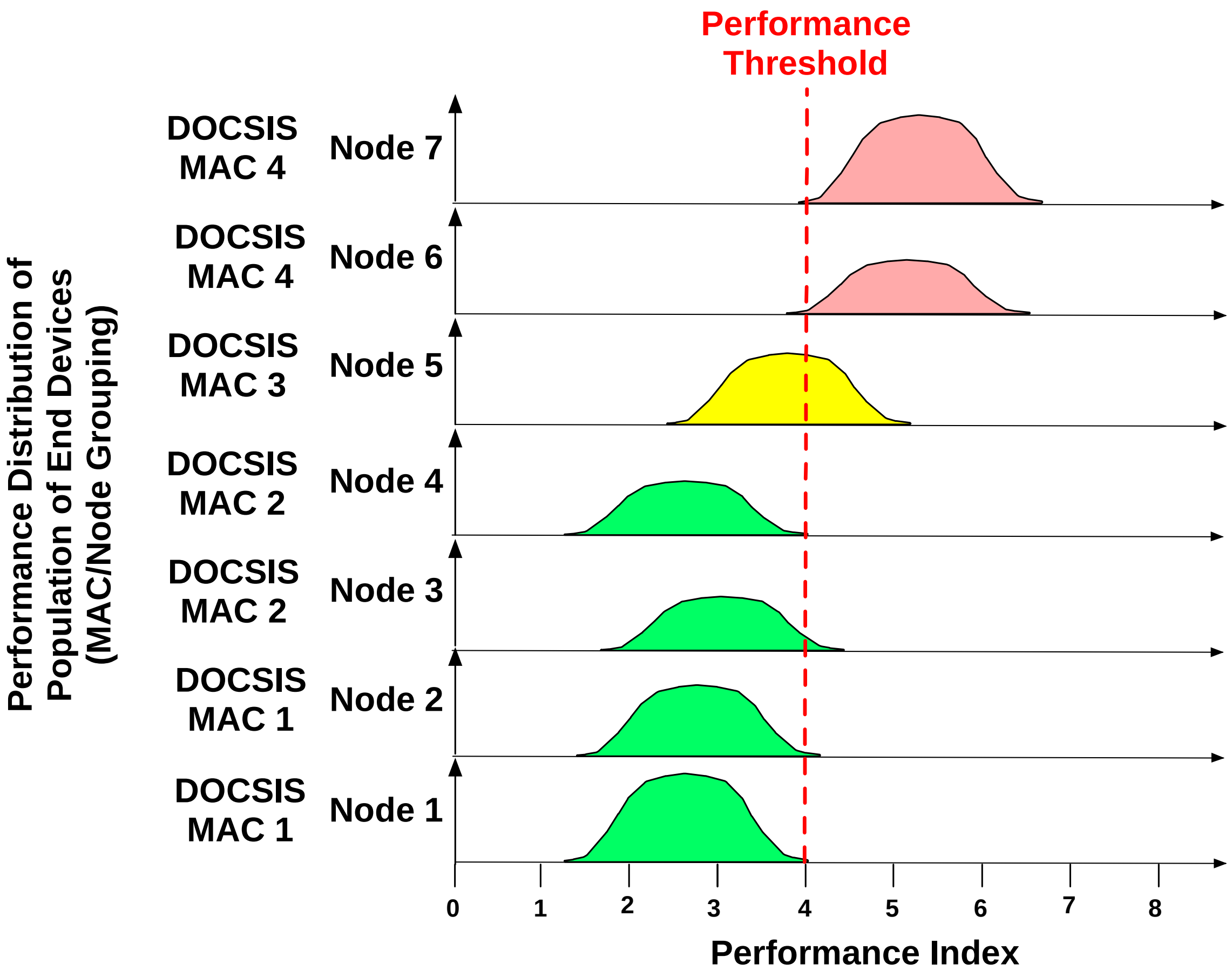

Figure 8 - Gaming and Video Streaming Performance Assessment

Figure 8 indicates that the latency problem extends across two fiber nodes, two isolated physical layers which share a common MAC layer domain. This symptom provides a high likelihood that this problem is a MAC or IP Layer problem. The analysis to troubleshoot and isolate the problem can be automated through relational database manipulation when performance metrics and topology information are stored in a granular database.

Figure 9 shows an example of how a hub network configuration. This configuration schematic highlights how the performance information from Figure 8 can be used to determine the source of the problem. A database with connectivity information across different layers with the associated performance data can be used to determine the source of the problem.

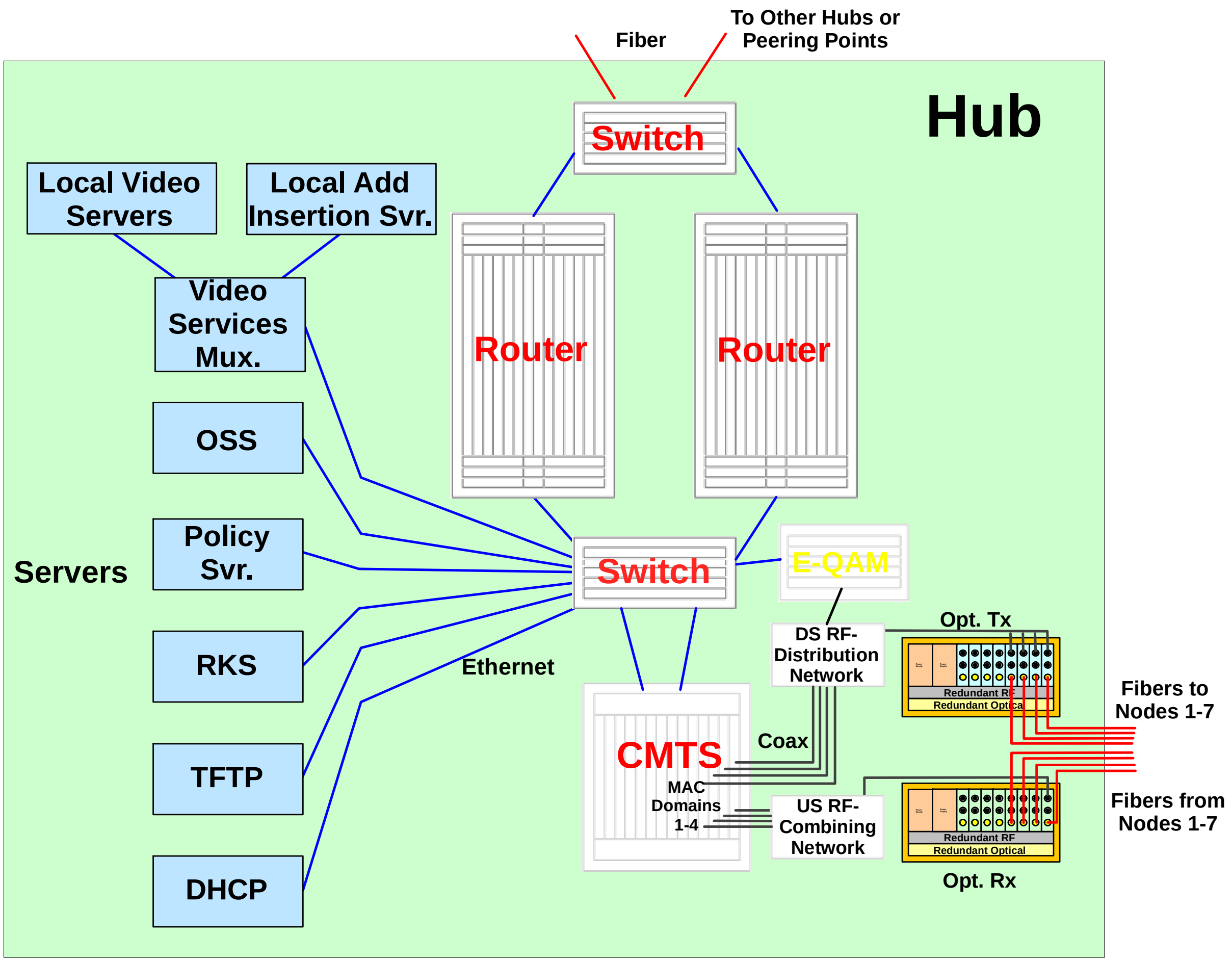

Figure 9 - Hub Video and Data Network Components

Subscriber service performance results of the network portion playing a role in delivering data services across different nodes within a Hub are shown in Figure 8. These performance results correspond to the specific Hub shown in Figure 9. The symptoms of nodes 6 and 7 are not observable in other Hubs of the regional aggregation network shown in Figure 10. In addition, traffic records indicate significant usage in nodes 6 and 7. This leads to believe that the cause of the problem is a layer 2 congestion issue within the MAC Layer domain shared by nodes 6 and 7.

The connectivity information of the regional network shown in Figure 10 could have been used to troubleshoot problems in the regional network. Many time when analyzing end-to-end services, the performance of the service has to be traced back from the local cable networks to the aggregation networks and potentially to the egress points of the cable network into other providers network and the public internet. It is important to discriminate the source and location of the problem of the problem so that action can be taken. In the case of problem responsibility lying outside the network this could range from communication to the third party network provider to reassessing peering arrangement that could address performance issues detected.

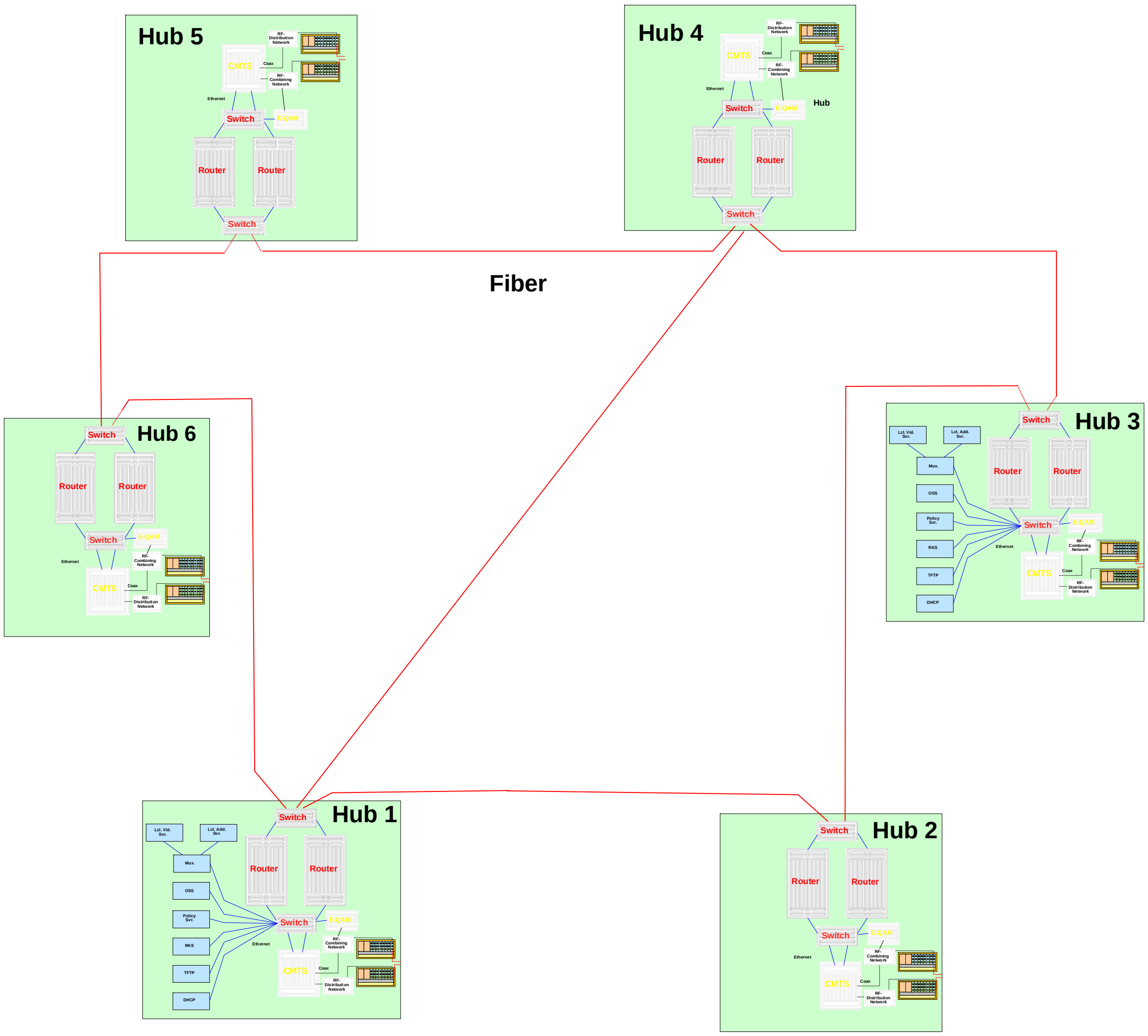

Figure 10 - Regional Cable Network

Second Use Case Scenario – Detection and Localization of Linear Distortions

Pre-equalization analysis has proven to be very effective at detecting and localization US impairments [3],[4]. This process involves analysis and correlation of data from CMs that share the same RF environment such as the same RF channel and optical node. The distortion signatures obtained from the CM pre-equalization coefficients are unique for each distortion within the node. Grouping the CMs that share the same impairment combined with topology information allows the determination of where the problem originates. Topology, hence device connectivity information as an integral part of the management databases becomes key at automating the fault location processes and determining impairment impact.

Figure 11 shows the linear distortion grouping results obtained from a node. This amplitude distortion versus frequency view of the channel by group of impacted CMs represent only a subset of the CMs with distortion.

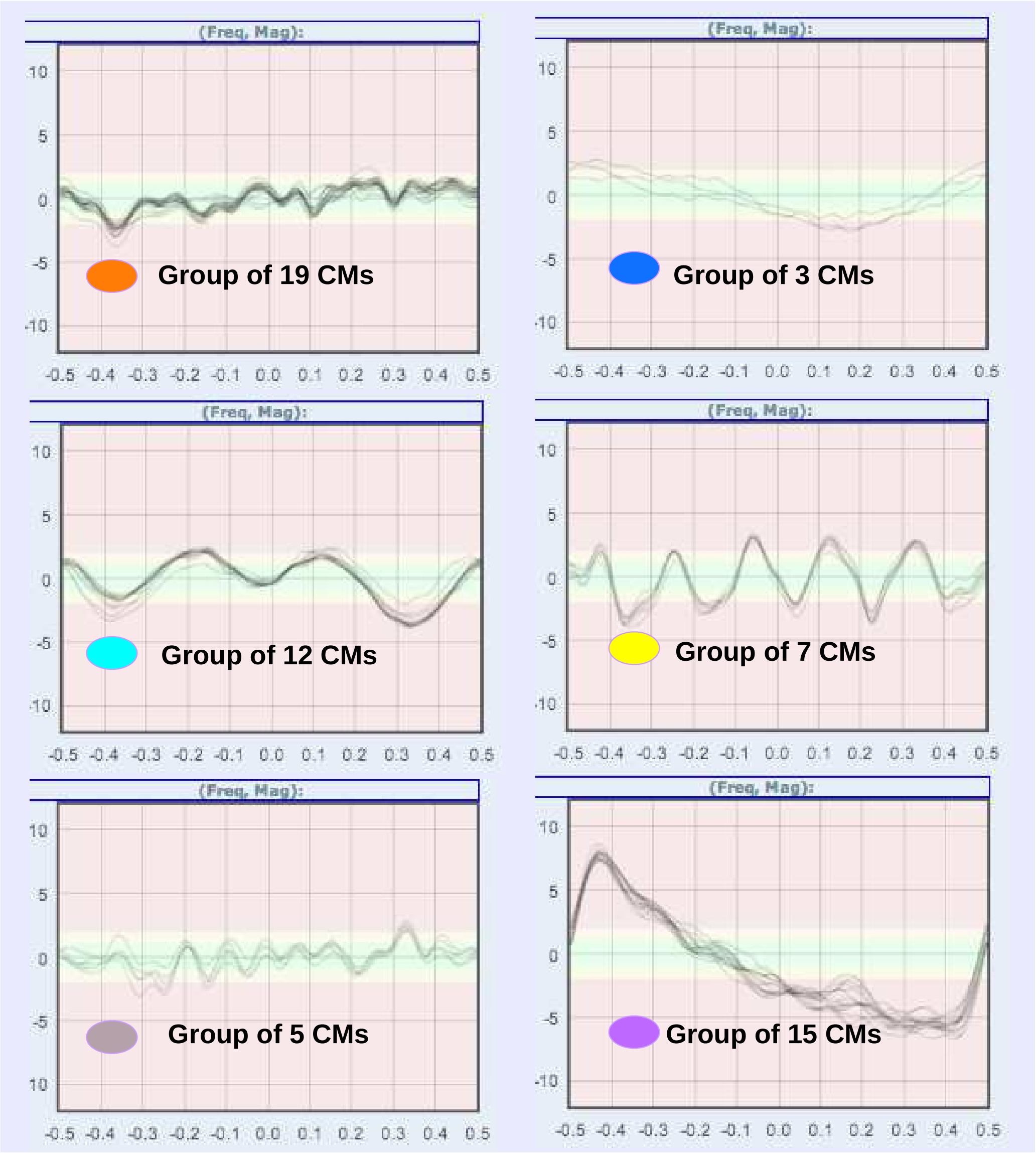

Figure 11 – Correlated Equalization Data from CMs in Node

Individual CMs with distortion are not shown in Figure 11 for brevity, but this determination is also important because that indicates an impairment in the drop/home portion of the network which should be addressed by a installer rather than a line technician.

Figure 12 shows a logical representation of an optical node with CMs that are color-coded based on the linear distortion grouping conducted for Figure 11. It is clearly seen that when the impairment correlation is combined with topology information, the determination of the location of the impairment is intuitive. It is the boundary (indicated by red Xs in Figure 12) between the topology regions that share the same impairment and the ones that don't.

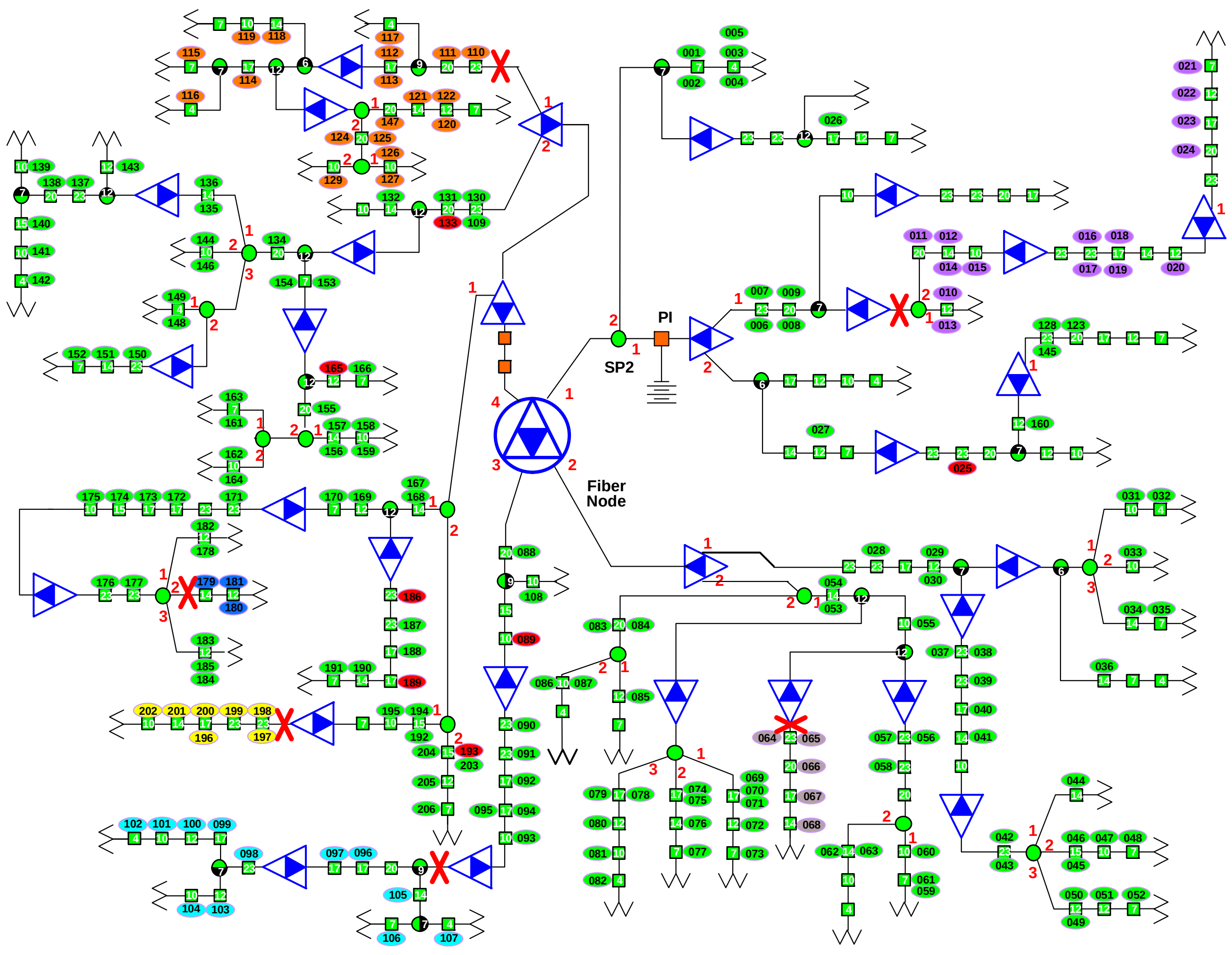

Figure 12 - Fiber Node Topology Representation Grouping Subscribers with Common Distortion

The color code used in Figure 12 is as follows; Green CMs have negligible distortion, Red CMs have measurable distortion but their distortion does not correlate with distortion of any other CM. Other colors indicate measurable distortion that correlates with the distortion of the CMs with the same color code.

Since this grouping technique also indicates how many CMs share the same impairment. Assessing severity by the number of CMs affected or type of customers affected can be implemented. Typically CM pre-equalization coefficients fully compensate for linear distortion impairments. When there is full compensation there is no other means for detecting that something is wrong with the network. This also facilitates a proactive network maintenance strategy since the pre-equalizers buy time for the operator to decide when to fix the problem. Incorporating distortion grouping information in a topology aware database can be used to automate the impairment discovery process.

## Third Use Case Scenario – Assessment of Resources to Optimize Node Splitting

Capacity in cable networks is typically measured with coarse granularity in time and by grouping individual user consumption on a node basis. An hour or 15 minute time granularity may hide through averaging short term capacity events. Likewise aggregate node consumption average out the traffic requirements few specific users may have. When traffic trends in a node project capacity starvation in a node, a node split is planned. This node split is typically done based on number of end devices covered.

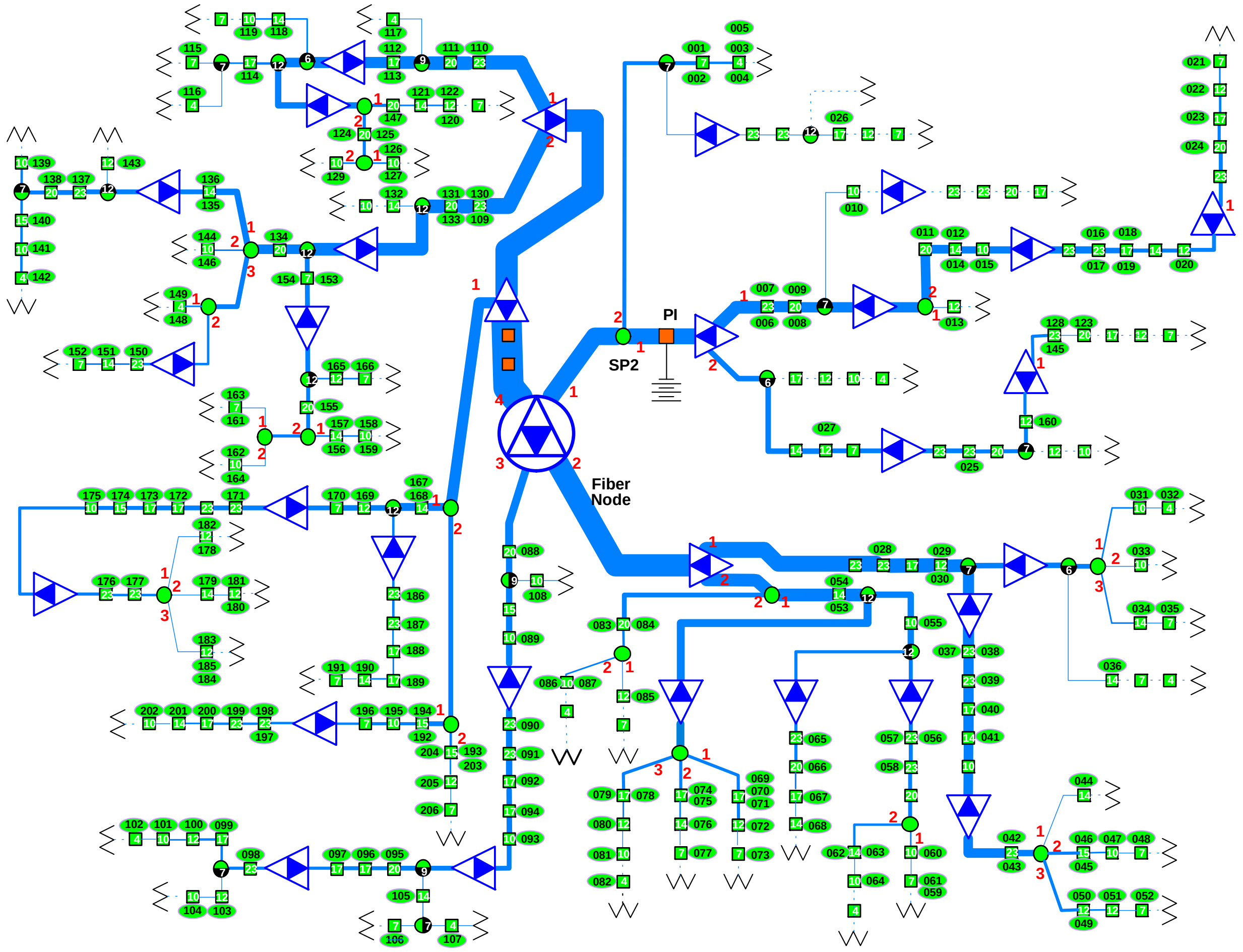

Figure 13 - Fiber Node Topology Representation Showing Subscriber Data Traffic Consumption

A different approach would be to have a node divided based on equal consumption. This can be implemented by monitoring end device consumption and correlating this information to physical network topology [4]. End device consumption information included as a data model element of the end device allows the automatic determination of traffic consumption for the nodes and any physical subset of the node.

Figure 13 highlights such an approach. In this case the individual consumption of users is aggregated as flows superimposed to topology to determine the actual consumption through each physical path. The time domain granularity increase the accuracy in determining that a channel is congested.

Service Availability and Reliability

A variety of use cases which benefit from granular network management and leverages connectivity and network layer awareness. In addition to impairment detection and localization, capacity planning, and node splitting applications, it can also be used for service availability, reliability and quality of experience assessment. In addition the insight in the network topology and the resources knowledge allows the determination of what type of business services can be provided and what SLAs can be met.

Statistical Process Control

Another potential application for granular data analysis can be derived leveraging the concepts of statistical process control promoted by Western Electric [5]. In this use-case the end-devices that are “out of control” or that not follow the expected behavior within 1, 2 or 3 standard deviations (sigmas) are identified within the topology and the same correlation exercises can be used to determine where within the network the problem originated.

Feedback Mechanisms

The use of subjective feedback from customers is very important. The increasing use of social networks as feedback tools can also be leveraged to determine customer satisfaction. Is that feedback concentrated in certain areas or is it generic to a system or a region? The same techniques to handle objective metrics and determine problem location can be also be used with subjective feedback.

The CM, STB and MTAs are quite intelligent and great probes to leverage for feedback on network health and to measure performance on services and applications. The most important characteristic of these devices is their ubiquity across the network. They are already in the customers home and can provide a wealth of information. This information can be included in the granular databases and used to manage the network. In many instances the cable industry has not taken advantage of many useful parameters available on these devices.

Operational Practices

The increasing amount of data gathered in the field requires also a change in the cable industry operational practices. We have to make sure that our databases reflect accurate information of our network. In particular for maintaining the HFC network, some manual processes are required. As part of the maintenance practices, cable technicians are changing and replacing components in the field. They are adjusting amplifiers, measuring plant performance and are exchanging end devices from customers. All these events have to be recorded to calculate reliability, estimate impact under impairments, to know the actual components in the network, to estimate resources such as available taps ports and available fiber, etc.

The technicians would have to enter greater amount of information to keep track of the changes in the level of detail that is envisioned. Technicians would also have to be empowered update the databases or through an automated verification process a technicians proposed changed would quickly validated by a supervisor.

Data Model Structure Standardization

Cable system operators have similar architectures, components, services they provide, operational practices, etc. Based on the similar environment and needs, it is advantageous to use similar databases using the same data model structure. The elements in the data model may vary from operator to operator but if it has the same structure, common management techniques can be implemented. In this paper several characteristics of a data model have been advocated. The main ones have been; granularity both in number of elements covered and in detail of description of each elements described, connectivity awareness, network layer awareness and end-to-end analysis friendliness. Standardization of a data model with these properties that is also extensible, is very useful for the cable industry. Standardization enables use of similar tools and processes and reduces the cost of implementation of management systems.

## CONCLUSION

This paper promoted a significant increase in the level of detail used to manage the cable network. It proposes to achieve a balance between reduction of data, by collapsing the different performance metrics into a few metrics and an increase in the number of elements and element parameters being managed. The analysis that is shown in DOCSIS® 3.0 indicates that two important performance metrics such as latency and packet loss are strongly correlated. Other performance metrics, in particular the ones used by the upper layers can be correlated with the fundamental lower layer performance metrics.

An in-depth knowledge of the network is also advocated. It is important to understand that the network is a fundamental part of the user experience. The amount of detail information proposed for collection along with the ubiquity of all cable end devices, provide significant insight into the network, the services and the customer. Control and management of the network enable management of the user experience. An example of a comprehensive granular data model has been shown. This data model is layered and is connection aware. Connectivity information is leveraged in use case scenarios for localizing linear-distortion impairments, estimating resources for node splitting applications and determining congestion points in the network. All of these use cases through the granular data model lend themselves to automation.

The unified management system approach presented here, which promotes breaking down the management system silos, facilitates end-to-end analysis. Leveraging a standard data model structure is fundamental for using similar tools across the industry and to leverage economies of scale.

The cable network has to be friendly to evolution and innovation, and for that it has to be open to leverage its resources for services and applications. A management system that is highly automated, comprehensive and granular enables that open environment.

## ACKNOWLEDGEMENT

The authors would like to thank Robert Cruickshank III for his encouragement and insightful feedback in writing this paper.